# Quantum States and Complex Projective Space

Bei Jia[1,2], Xi-guo Lee[1,3]

[1]*Institute of Modern Physics, Chinese Academy of Sciences, P.O.Box 31 Lanzhou, 730000, China*

[2]*Graduate School of Chinese Academy of Sciences, Beijing, 100080, China*

[3]*Center of Theoretical Nuclear Physics, National Laboratory of Heavy Ion Collisions,*

*P.O.Box 31 Lanzhou, 730000, China*

## Abstract

In this paper we propose the idea that there is a corresponding relation between quantum states and points of the complex projective space, given that the number of dimensions of the Hilbert space is finite. We check this idea through analyzing some of the basic principles and concepts of quantum mechanics, including the principle of superposition, representations and inner product of quantum states, and give some interesting examples. Based on our point of views we are able to generate the evolution equation of quantum states ---- the Heisenberg equation. We also discuss the act of dynamical operators on quantum states.

## 1  Introduction: Complex Projective Space and Quantum States

It has been considered that there exists a geometry description of quantum mechanics other that the conventional description in Hilbert space [1] [2] [3] [4] [5] [6]. In these articles it is argued that the quantum state space (or quantum phase space) is a projective Hilbert space. Here we present the idea that the state space can be viewed as a complex projective space if the number of dimensions of the Hilbert space is finite, and from this point of view we are able to get some principles and dynamical features of conventional quantum mechanics. A similar idea can be found in [5] [6]. Here we examine this proposal in detail and treat it in some different ways.

A complex projective space $CP^n$ is defined as follow [7]: it is the set of all the lines that pass through the origin of the $C^{n+1}$ space, with an equivalent relation that every single line is regarded as a point in this multiset. That is to say, we have an equivalent relation $z \sim cz$ in the $C^{n+1}$ space

$$(z^0, z^1, \cdots, z^n) \sim (cz^0, cz^1, \cdots, cz^n), \forall c \in C - \{0\} \tag{1.1}$$

These $z^0, z^1, \cdots, z^n$ cannot be zero simultaneously. We can write this equivalent relation as

$$[z] = [z^0, z^1, \cdots z^n] \tag{1.2}$$

Every $[z]$ represents a point in $CP^n$, that is to say, every point in $CP^n$ has a set of homogeneous coordinates $[z^0, z^1, \cdots z^n]$.

Some features of $CP^n$ [7]. Points in $CP^n$ are equivalent to points in the spherical surface $S^{2n+1}$ of $C^{n+1}$ space, multiplied with a phase factor of $\{Z^k\} \sim \{e^{i\alpha} Z^k\}$. Therefore,

$$CP^n \cong S^{2n+1}/U(1) \cong C^n \bigcup CP^{n-1} \tag{1.3}$$

On the other hand, the transitive group of $CP^n$ is $SU(n+1)$, that is to say, for every $g \in SU(n+1)$, $p, q \in CP^n$, there is a map

$$\begin{aligned}\varphi_g : SU(n+1) \times CP^n &\to CP^n \\ p &\mapsto \varphi_g(p) = q\end{aligned} \quad (1.4)$$

The isotropic group of $CP^n$ is the $U(n)$ group, hence

$$CP^n \cong SU(n+1)/U(n) \quad (1.5)$$

According to the basic principles of quantum mechanics, all the eigenstates of a dynamical operator can form a complete set of basis vectors of a Hilbert space, and any quantum state of a quantum system can be regarded as a vector in this Hilbert space. For example, we have a dynamical operator $A$, with its eigenequation $A|k\rangle = A'_k|k\rangle$. $A'_k$ is the eigenvalue of $A$, and $|k\rangle$ is the corresponding eigenstate. Then an arbitrary quantum state of a quantum system can be written as

$$|\psi\rangle = \sum_k \langle k|\psi\rangle|k\rangle = \sum_k a_k|k\rangle \quad (1.6)$$

where $a_k = \langle k|\psi\rangle$. We restrict our discussion to the condition that these eigenstates are discrete. We also do not refer to normalization here because our purpose is to analyze the most basic situation. We can see the quantum state $|\psi\rangle$ becomes a vector in a Hilbert space, whose basis vectors are the $|k\rangle$s. That is to say, every quantum state $|\psi\rangle$ has a corresponding vector in the Hilbert space, whose coordinates are $(a_1, a_2 \cdots, a_k, \cdots)$. Usually speaking the number of the dimensions of this space is infinite, but we only discuss the condition of finite dimensions here. We know that a state is specified by the direction of the vector and any length one may assign to the vector is irrelevant. That is to say, if we multiply the vector with an arbitrary complex number $c$, the resulting state must be the same as the original one apart from the special case that $c$ equals to zero, so that the vector $c|\psi\rangle$ must correspond to the same state as $|\psi\rangle$ does [8]. Therefore we can define an equivalent relation

$$(a_1, a_2, \cdots, a_{n+1}) \sim (ca_1, ca_2, \cdots, ca_{n+1}), \forall c \in C - \{0\} \quad (1.7)$$

and write this equivalent relation as

$$[a] = [a_1, a_2, \cdots, a_{n+1}] \quad (1.8)$$

If we look back at the definition of $CP^n$, we will realize that we actually defined the homogeneous coordinates of a point in this $CP^n$. Thus a quantum state $|\psi\rangle$ actually corresponds to a point in $CP^n$, with its homogeneous coordinates being $[a_1, a_2, \cdots a_{n+1}]$.

It is import to notice that the complex numbers $z^0, z^1, \cdots, z^n$ in the definition of $CP^n$ cannot be zero simultaneously. For the state $|\psi\rangle$, if $a_1, a_2, \cdots, a_{n+1}$ all equal to zero, then $|\psi\rangle = 0$, and this does not make any sense. Hence the demand from the definition of $CP^n$ that $a_1, a_2, \cdots, a_{n+1}$ cannot be zero simultaneously is satisfied automatically.

## 2  A Simple Example: The Spin of an Electron

In order to vividly understand the argument above, we can refer to the example of electron spin. There is a similar example about spin 1 particle in [6]. Here we write the spin eigenstates of a single electron as $|1/2\rangle$ and $|-1/2\rangle$, then an arbitrary spin state of the electron $|\psi\rangle$ can be written as

$$|\psi\rangle = a|1/2\rangle + b|-1/2\rangle \tag{2.1}$$

where $a = \langle 1/2|\psi\rangle$, $b = \langle -1/2|\psi\rangle$.

According to the discussion in the first section, we can say that this quantum state $|\psi\rangle$ has a corresponding vector in the 2-dimensional complex linear space, whose coordinates are $(a,b)$. We know that the length of $|\psi\rangle$ as a vector is not important, so we can define an equivalent relation

$$(a,b) \sim (ca, cb), \quad \forall c \in C - \{0\} \tag{2.2}$$

That is also to say, we can give this vector a set of homogeneous coordinates $[a,b] = [ca, cb]$, so that $|\psi\rangle$ has a corresponding point in $CP^1$. To sum up, we can say that every spin state of an electron has a corresponding point in $CP^1$.

We know from group theory that spin is related to the $SU(2)$ group [9]. In the first section we also mentioned that $CP^n \cong SU(n+1)/U(n)$, so $CP^1 \cong SU(2)/U(1)$. This is another explanation to the relation between spin and $SU(2)$ group.

## 3 The Principal of Superposition of States

A similar but different discussion about superposition can be found in [6]. Our purpose here is to check the validity of this principle in detail. According to one of the basic principals in quantum mechanics, the principal of superposition of states, if $|A\rangle, |B\rangle, |C\rangle, \cdots$ are all possible quantum states of a quantum system, then

$$|\psi\rangle = c_1|A\rangle + c_2|B\rangle + c_3|C\rangle + \cdots \tag{3.1}$$

is also a possible quantum state of this system, where $c_1, c_2, c_3, \cdots$ are arbitrary complex numbers which cannot be zero simultaneously.

We know from above that a quantum state $|\psi\rangle$ has a corresponding point in $CP^n$, this means that if

$$[z] = c_1[z_1] + c_1[z_2] + c_1[z_3] + \cdots \tag{3.2}$$

where $[z_1], [z_2], [z_3] \cdots$ are all points in $CP^n$, and $c_1, c_2, c_3, \cdots$ are arbitrary complex numbers which cannot be zero simultaneously, then $[z]$ should also be a point in $CP^n$. This is obviously true since there is an equivalent relation (1.1) in the definition of $CP^n$. Hence the parallelism between quantum states and points in $CP^n$ does not violate the principal of superposition of states.

There is a special case that we should put attention to. If $|\psi\rangle$ is a quantum state, then

$$|\psi'\rangle = c_1|\psi\rangle + c_2|\psi\rangle + c_3|\psi\rangle + \cdots = (c_1 + c_2 + c_3 + \cdots)|\psi\rangle \tag{3.3}$$

is not a different state from $|\psi\rangle$ but the same state as $|\psi\rangle$, where $c_1, c_2, c_3, \cdots$ are arbitrary complex numbers which cannot be zero simultaneously and $c_1 + c_2 + c_3 + \cdots \neq 0$. That is to say, if the vector corresponding to a state is multiplied by any complex number, not zero, the resulting vector will correspond to the same state [8]. If we look back at the definition of $CP^n$, we can see this is exactly the starting point of our argument.

## 4 Representations and the Inner Product of Quantum States

In the discussion above we came to the conclusion that every quantum state has a corresponding point in $CP^n$. The value of $n$ depends on the basis vectors ---- the eigenstates that we choose. Based on this point we can say the local coordinates of a point in $CP^n$ correspond to the projections of the quantum state, which correspond to this point, on the eigenstates that we choose ---- the representation. That is to say, to choose a set of local coordinates of a point in $CP^n$ is equivalent to choose a set of eigenstates, then the local coordinates are the representation of the quantum state which corresponds to this point. For example, we have two dynamical operators $A$ and $B$, whose eigenstates are $|k\rangle$ and $|l\rangle$ respectively. We have to restrict to the condition that the number of eigenstates of $A$ and $B$ are the same, say $n+1$, then an arbitrary state $|\psi\rangle$ can be written as

$$|\psi\rangle = \sum_k \langle k|\psi\rangle |k\rangle = \sum_k a_k |k\rangle = \sum_l \langle l|\psi\rangle |l\rangle = \sum_l b_l |l\rangle \tag{4.1}$$

According to the principles of quantum mechanics, column matrices $(a_1, a_2, a_3, \cdots)^T$ and $(b_1, b_2, b_3, \cdots)^T$ are the representations of $|\psi\rangle$ under $A$ and $B$ respectively. On the other hand, $[a_1, a_2, a_3 \cdots]$ and $[b_1, b_2, b_3 \cdots]$ are exactly two sets of local coordinates of the point which corresponds to the state $|\psi\rangle$. Hence the local coordinates represent the representations, and the switch between different representations is equivalent to the switch between different local coordinates in $CP^n$.

We can write this point of view in the language of map. State $|\psi\rangle$ corresponds to a point $p \in CP^n$, then to choose the $A$ representation means to have a map $\varphi$

$$\begin{aligned} \varphi : CP^n &\to C^n \\ p &\mapsto \varphi(p) = (a_1, a_2, a_3, \cdots) \end{aligned} \tag{4.2}$$

Similarly, to choose the $B$ representation is equivalent to a map $\phi$

$$\begin{aligned} \phi : CP^n &\to C^n \\ p &\mapsto \phi(p) = (b_1, b_2, b_3, \cdots) \end{aligned} \tag{4.3}$$

We write $(a_1, a_2, a_3, \cdots)$ as $\mathbf{a}$, and $(b_1, b_2, b_3, \cdots)$ as $\mathbf{b}$ for simplicity. Then the switch from $A$ representation to $B$ representation is

$$\mathbf{b} = \phi(\varphi^{-1}(\mathbf{a})) \tag{4.4}$$

There is another critical point that needs to be considered. We all know that a quantum state can be expressed as a vector in Hilbert space, and a Hilbert space is a vector space with complete inner product. Therefore we need to add the concept of inner product into our discussion. The inner product of two quantum states is defined as

$$\left(|\psi\rangle, |\psi'\rangle\right) \equiv \langle \psi'|\psi\rangle \tag{4.5}$$

This can be written under $A$ representation as

$$\langle \psi'|\psi\rangle = \sum_{k'}\sum_k \langle k'|a_{k'}'^* a_k|k\rangle = \sum_{k'}\sum_k a_{k'}'^* a_k \delta_{k'k} = \sum_k a_k'^* a_k \tag{4.6}$$

If states $|\psi\rangle$ and $|\psi'\rangle$ correspond to points $p$ and $p'$ in $CP^n$, then $\mathbf{a}$ and $\mathbf{a}'$ are the local coordinates of $p$ and $p'$ respectively. Thus the inner product can be expressed as a map

$f$

$$\begin{aligned} f: CP^n &\to C \\ p, p' &\mapsto f(p, p') = \sum_k a'^*_k a_k \end{aligned} \quad (4.7)$$

## 5 Evolution Equation and Tangent Bundle

Based on the discussion above, now we can go on to analyze the evolution equation of quantum states ---- the Schrödinger equation. This problem is also discussed in [2] [6], but in different aspects.

Let's say we have a point $p \in CP^n$, whose corresponding quantum state is $|\psi\rangle$. Under $A$ representation $|\psi\rangle$ can be expressed as a column matrix $(a^1, a^2, a^3, \cdots)^T$, then the matrix form of Schrödinger equation is [10]

$$i\hbar \frac{d}{dt} a^i(t) = \sum_j H_{ij} a^j \quad (5.1)$$

On the other hand, we can define tangent vector on $CP^n$ since it is a complex differential manifold. We choose the local coordinates of $p$ as $(a^1, a^2, a^3, \cdots)$, then the tangent vector on the point $p$ is

$$X_p = \sum_i \frac{da^i}{dt} \frac{\partial}{\partial a^i} \quad (5.2)$$

where $t$ is an arbitrary parameter, and $da^i/dt$ is the component of $X_p$ when we choose the local coordinates as $(a^1, a^2, a^3, \cdots)$ [7]. If we compare (5.1) and (5.2), we will realize that the components of the tangent vector on $CP^n$, the $da^i/dt$ s, are actually the linear combination of the local coordinates on the base manifold, and the coefficients are the elements of the matrix form of the Hamiltonian operator $H$, if we choose parameter $t$ as time. In another word, the act of the Hamiltonian operator $H$ on quantum state can be regarded as a map from $CP^n$ to the tangent space $T_p(CP^n)$ on it. By saying this we mean that this map actually gives every point $p \in CP^n$ a corresponding tangent vector $X_p \in T_p(CP^n)$. As a whole we actually find a map from $CP^n$ to the tangent bundle on it, from which we give a tangent vector field $X_p$.

There is another important aspect of this problem, and we will discuss it in the Heisenberg picture. The evolution equation of a dynamical operator $L$ in the Heisenberg picture is the Heisenberg equation

$$\frac{d}{dt} L(t) = \frac{1}{i\hbar} [L(t), H] \quad (5.3)$$

where $H$ is the Hamiltonian operator. In quantum mechanics a dynamical operator is actually a map between two states, which mapping one state to another [8]. Therefore a dynamical operator can be regarded as a function defined on states, and according to our discussion it is also a function defined on the point of $CP^n$.

Let's turn our focus into the mathematical part. A $CP^n$ space is actually a symplectic manifold, so we can define Hamilton vector field $X_H$ on $CP^n$. Then the single parameter variation of a function $L(p)$ on $CP^n$ along $X_H$ is

$$\frac{d}{dt}L(p) = X_H L = \{L, H\} \tag{5.4}$$

where $t$ is the parameter, $\{\}$ is the Poisson bracket, and $H$ is the Hamiltonian function defined on $CP^n$ [7]. In quantum mechanics we have the canonical quantization that $\{L, H\} = [L, H]/i\hbar$, as long as $L(p)$ is a dynamical operator and $H$ is the Hamiltonian operator. If we choose this parameter $t$ as time, then equation (5.4) is exactly like the Heisenberg equation, except it is defined on the $CP^n$ space rather that on Hilbert space. This is actually the evolution equation of quantum states based on the symplectic structure of $CP^n$.

## 6 Dynamical Operator

We have used the concept of dynamical operator since the beginning of this paper. Now we need to discuss what an operator really dose to quantum states in our system. Let's say we have an operator $A$ whose eigenequation is $A|k\rangle = A_k'|k\rangle$, where $A_k'$ is the eigenvalue of $A$, and $|k\rangle$ is the corresponding eigenstate, then we have equation (1.7). Therefore if we act this operator on $|\psi\rangle$, we will have

$$A|\psi\rangle = \sum_k a_k A|k\rangle = \sum_k a_k A_k'|k\rangle = \sum_k a_k'|k\rangle = |\psi'\rangle \tag{6.2}$$

where $a_k' = a_k A_k'$. $|\psi'\rangle$ is also a possible quantum state according to the principal of superposition of states. Hence if we act an operator on a state, we will get another state. In fact in section four we mentioned that in quantum mechanics a dynamical operator is actually a map between two states, which mapping one state to another [8].

From the view of differential geometry, if points $p \in CP^n$ and $p' \in CP^n$ corresponding to states $|\psi\rangle$ and $|\psi'\rangle$ respectively, then the act of $A$ on $|\psi\rangle$ can be regarded as a map from $p$ to $p'$

$$A: CP^n \to CP^n$$
$$p \mapsto A(p) = p'$$

(6.3)

Generally speaking, dynamical operator $A$ is required to be a Hermitian operator, so the matrix forms of $A$ should be Hermitian matrices, which can be expanded a set of basis of traceless matrices. We know that the transitive group of $CP^n$ is $SU(n+1)$, and the generator of the $SU(n+1)$ group are some traceless matrices, so there might be some relation between dynamical operators and the transitive group $SU(n+1)$.

There is an interesting fact here. In the second section we mentioned that every spin state of an electron has a corresponding point in $CP^1$, then the act of the spin operator $\mathbf{s}$ on a spin state $|\psi\rangle$, which generate another spin state $|\psi'\rangle$, is in fact a map

$$\mathbf{s}: CP^1 \to CP^1$$
$$p \mapsto \mathbf{s}(p) = p' \tag{6.4}$$

where $p$ and $p'$ are points in $CP^1$ and corresponding to states $|\psi\rangle$ and $|\psi'\rangle$ respectively. It is important to notice that $\mathbf{s} = \frac{\hbar}{2}\boldsymbol{\sigma}$, where $\boldsymbol{\sigma}$ is the Pauli operator, and the Pauli matrices $\sigma_1, \sigma_2, \sigma_3$ are the generators of the self-representation of the $SU(2)$ group, which is the transitive group of $CP^1$.

# 7 Conclusion

Relationship between quantum states and points of the complex projective space is examined in this article. We check the validity of several principles and features of conventional quantum mechanics, and some dynamical problems are discussed. But we only considered the finite dimensional case. In most cases of quantum mechanics the number of dimensions of the Hilbert space is infinite, so a generalization of this idea to the infinite dimensional case is needed. There is another important problem that needs to be clarified, that what the relationship really is between dynamical operators and the transitive group of complex projective space.